\documentstyle[aps,preprint]{revtex}

\widetext
\draft
\tighten
\begin{document}

\preprint{EFUAZ FT-94-08-REV}

\title{Speculations on the Neutrino Theory of Light.\thanks{Submitted to
``Physics Essays". A part of this work has been reported at the
IFUNAM seminar, M\'exico, 2/IX/1994.}}

\author{{\bf Valeri V. Dvoeglazov}\thanks{On leave of absence from
{\it Dept. Theor. \& Nucl. Phys., Saratov State University,
Astrakhanskaya ul., 83, Saratov\, RUSSIA.} Internet
address: dvoeglazov@main1.jinr.dubna.su}}

\address{
Escuela de F\'{\i}sica, Universidad Aut\'onoma de Zacatecas \\
Apartado Postal C-580, Zacatecas 98068, ZAC., M\'exico\\
Internet address:  valeri@cantera.reduaz.mx\\
URL: http://cantera.reduaz.mx/\~~valeri/valeri.htm
}

\date{February, 1998. First version: September, 1994}

\maketitle

\begin{abstract}
The basis of new ideas in the old theory is the Majorana and Ahluwalia
constructs, modified versions of the Weinberg $2(2j+1)$ theory,
and the Barut's self-field quantum electrodynamics.
\end{abstract}


\newpage

\section{Answers}

The neutrino theory of light has a long
history~\cite{DeBroglie}-\cite{Ray}.  Its crucial point  was the Pryce
theorem~\cite{Pryce} which  brought a halt in the development of this
branch of particle physics for a long time.
Namely, as showed Barbour {\it et al.} in the most elegant way,
ref.~\cite{Barbour}, the Jordan's ``ansatz"\footnote{${\cal
A}_\mu (k)$ is the electromagnetic potential in the momentum space. Here
$k$ is the photon four-momentum, $\gamma_\mu$ the usual $4\times 4$
matrices and $\psi$ is the spin-1/2 field operator.  $f (\lambda)$ is some
suitable weight function.  In the de Broglie theory one has $f (\lambda) =
\delta (\lambda -a)$.}
\begin{equation}\label{ansatz} {\cal  A}_{\mu} (k)
= \int_0^1 f (\lambda) \bar \psi ((\lambda -1) k) \gamma_{\mu} \psi
(\lambda k) d\lambda
\end{equation}
was thought to be incompatible with the requirement
of  the transversality of a photon and/or with the requirements
of the correct statistics for a photon.

In the sixties some hopes on recreation emerged, the renaissance occurred.
Berezinsky writes~\cite{Berezinsky}: ``With a gauge transformation
the amplitude of a photon state of the type:
\begin{equation}
m\bar \psi (\tau k)\gamma_\mu \psi (\lambda k) = -ik_\mu \bar
\psi (\tau k) \psi (\lambda k)
\end{equation}
($\lambda$ and $\tau$ are numbers)
can always be reduced to zero" and then he disagrees with I. M. Barbour
{\it at al.}: ``In the four-component theory there is
no difficulty in practically constructing all transverse
four-vectors from the wave functions of neutrinos with
collinear momenta".

Sen, ref.~\cite{Sen}, writes: `` \ldots the 2-component neutrino can also
be completely described by a self-dual antisymmetric tensor
behaving very much like the
electromagnetic field tensor.  What about the anti-self-dual  case?\ldots
We have shown that the photon can be considered as a combination of
two neutrinos, one described by
\begin{equation}\label{n1}
(\sigma_k \partial_k +i\partial_4 )\phi=0
\end{equation}
and other by
\begin{equation}\label{n2}
(\sigma_k^\prime \partial_k +i\partial_4 )\chi =0
\end{equation}
(the set $\sigma_k^\prime$ differs from $\sigma_k$ only by
the interchange of the indices one and three) and that  the neutrinos
described by (\ref{n1}) and (\ref{n2}) can be represented by a self-dual
and an anti-self-dual antisymmetric tensor, respectively.
The two neutrinos will behave
identically when they are free but their interactions with other particles
may be different. The problem of how they differ in their
interactions remains to be investigated."

Bandyopadhyay,  ref.~\cite{Bandy3},
continues: ``Perkins, ref.~\cite{Perkins}, considered the usual
four-component solutions with definite momentum and helicity
of the Dirac equation
with a zero-mass term. He constructed in some special
way the electromagnetic field tensor
$F_{\mu\nu}$ of the two neutrino operators in the
configurational space. In this
formalism, the photon appears as composed of the pair $(\nu_1 \bar \nu_2)$
and $(\nu_2 \bar \nu_1)$, where $\nu_1$ $(\nu_2)$ denotes the neutrino
with spin parallel (antiparallel)
to its momentum. However, the impossibility of constructing linearly
polarized photons seems to be a very serious defect of this theory.\ldots
we can say that the construction of the photon
as a composite state of a neutrino-antineutrino
pair has remained a problem until now."

Strazhev concludes~\cite{Strazhev}: ``Neutrino and photon states realize
the IR [irreducible representations] of chiral group $U(1)$
and all the relations that connect the neutrino and photon
operators must have definite
transformation properties under chiral transformation.
The electromagnetic potentials
${\cal A}_\mu  (k)$ are not dual invariants. At the same
time the right side of the expression
(\ref{ansatz}) is an invariant of $\gamma_5$ transformations. Of course, from
the particular solutions of the Dirac equation with mass $m=0$
can be built the transversal
four-vector (Berezinsky, 1966, ref.~\cite{Berezinsky}) \ldots In this
case, however, the condition of the relativistic invariance will not be
satisfied. So, we can formulate the Pryce theorem in the following way:
The requirements of the correct statistics and the correct transformation
properties under transformation of the group $U(1)$ of the
composite photons are incompatible in
the neutrino theory of light. In other words we
can say that with the satisfaction
of the requirement of the relativistic and chiral invariance
one can build from
the neutrino only a photon with unphysical longitudinal
polarization."                .

A succeeding halt! \ldots

\section{Thoughts}

I would still like to mention here: very unfortunately the paper of
Sen~\cite{Sen} did not draw  almost  any attention.  I believe that  it
was in the needed direction. Furthermore, it is not clear to me, why do
peoples, speaking about the {\it neutrino} theory of light, used the Dirac
field operator which describes {\it charged particles}? Finally, in
the recent textbook on the quantum field theory~\cite{TEXTBOOK} S.
Weinberg indicated the possibility that the 4-vector potential can be used
to describe a scalar particle\ldots  So, it seems to me that some
speculations about the problem, which is overweighed by many confusions,
would be desirable.

In ref.~\cite{Berezinsky} the following assumptions have been made,
under which one can have   the consistent  neutrino theory of light.

``{\it 1. Let $U_\theta$ be an operator which in the space of neutrino state
vectors describes a rotation by angle $\theta$ around the direction of
the momentum ${\bf k}$ as an axis. Then}
\begin{equation}
U_\theta \, a_i ({\bf k})\, U_\theta^{-1} \,= \, e^{i\theta/2}\,
a_i ({\bf k})\quad, \quad
U_\theta \, b_i ({\bf k})\, U_\theta^{-1} \,=\, e^{-i\theta/2}\,
b_i ({\bf k})\quad .
\end{equation}

{\it 2. If the neutrinos are fermions, then the operators $a_i ({\bf k})$
and $b_i ({\bf k})$
obey the following commutation relations:}
\begin{equation}
\left [a_i^+ ({\bf k}), a_j ({\bf k}^{\,\prime}) \right ]_+ =
\left [ b_i^+ ({\bf k}), b_j ({\bf k}^{\,\prime}) \right ]_{+} =
\delta_{ij} \delta ({\bf k} - {\bf k}^{\,\prime})\quad.
\end{equation}
{\it All other anticommutators are equal to zero.}\footnote{We still note
that another set of anticommutation relations has been recently
proposed~\cite{DVA2} in the Majorana-McLennan-Case-like construct.}

{\it 3. There exist photons with right and left circular polarization,
whose annihilation
operators $\kappa ({\bf p})$ and $\omega ({\bf p})$ have
the following transformation
properties:}
\begin{equation}
U_\theta \,\kappa ({\bf p})\, U_\theta^{-1} \,= \,e^{i\theta}\,
\kappa ({\bf p})\quad,\quad
U_\theta \,  \omega ({\bf p})\, U_\theta^{-1}\, = \, e^{-i\theta}\,
\omega ({\bf p})\quad .
\end{equation}

{\it 4. Simultaneously there exist photons with linear polarization, whose
annihilation operators $\xi ({\bf p})$ and $\eta ({\bf p})$ are
linear combinations of
$\kappa ({\bf p})$ and $\omega ({\bf p})$ and satisfy
the transformation relations}
\begin{mathletters}
\begin{eqnarray}
U_\theta \,\xi ({\bf p}) \,U_\theta^{-1} \,&=&\, \xi ({\bf p})\,
\cos \theta \,+\,\eta ({\bf p}) \sin \theta\quad,\\
U_\theta \, \eta ({\bf p}) \, U_\theta^{-1}\, &=&\, -\,\xi ({\bf p})\,
\sin \theta \,+\,\eta ({\bf p}) \, \cos \theta\quad .
\end{eqnarray}
\end{mathletters}

{\it 5. It is also assumed that the photon operators satisfy
the following commutation relations:}
\begin{equation}
\left [ \xi ({\bf p}), \eta^+ ({\bf p}) \right ]_{-} =0, \quad
\left [ \kappa ({\bf p}), \omega^+ ({\bf p}) \right ]_{-} =0\quad .
\label{cr}
\end{equation}
{\it Equations (\ref{cr}) are the conditions for the photon
to be genuinely neutral. No assumptions
are made about the equations for the photons or the neutrinos."}

Strazhev, ref.~\cite{Strazhev},
reformulated them in a following way: `` The operators of
the massless fields have the definite transformation rules
under chiral ($\gamma_5$) transformations;
there exist field operators $\vec E$, $\vec H$ that are transformed
under dual transformations in accordance with
\begin{eqnarray}
U_\theta \,\Phi_i \,U_\theta^{-1} \,&=&\, \Phi_i \,\cos\theta \,+
\,\tilde \Phi_i \sin\theta\\
U_\theta \,\tilde \Phi_i \,U_\theta^{-1}\, &=&\, -\,\Phi_i \, \sin\theta \,+
\,\tilde \Phi_i \, \cos \theta\quad,
\end{eqnarray}
(where $i=1,2$); a neutrino can be either a fermion or a parafermion particle;
the commutation relations of the operators of
the photon fields are invariants
under dual transformations\ldots The last condition is an equivalent
to the condition of the pure neutrality
of the photon by Berezinsky\ldots We do not
have a self-consistent neutrino theory
of light if all these conditions are satisfied. From
our point of view, that means that the statistical
properties of the photon in
the neutrino theory of light are inconsistent with
the chiral ($\gamma_5$) symmetry of neutrino and electromagnetic fields."

\smallskip

At this point I would like to remind recent results obtained in the
framework of the $2(2j+1)$ component theory~\cite{Weinberg}-\cite{Tucker},
the antisymmetric tensor field description~\cite{Hayashi,Kalb} and in the
Majorana theory of neutral particles~\cite{Majorana,CaseM}.

\begin{itemize}

\item
D. V. Ahluwalia and D. J. Ernst found that the Weinberg
first-order equations for massless free particle of arbitrary spin
\begin{mathletters}
\begin{eqnarray}
\left ({\bf J} \cdot {\bf p} - j p^0\right ) \phi_{_R} ({\bf p}) &=&
0\quad, \label{Max1}\\
\left ({\bf J} \cdot {\bf p} + j p^0\right ) \phi_{_L}
({\bf p}) &=& 0\label{Max2}
\end{eqnarray}
\end{mathletters}
have acausal solutions. For instance, for the spin $j=1$ case one has
the puzzled solution $E=0$, see Table 2 in ref.~\cite{Ahl0}.
The satisfactory explanation is required to this solution not only from
the physical viewpoint but from the methodological viewpoint
too.\footnote{See also the discussion in refs.~\cite{D40,D4}.
The concept of `action-at-a-distance' presented by A. E. Chubykalo and R.
Smirnov-Rueda~\cite{CHUB} may also be  connected with the problem at
hand.  The situation is now similar to that which we
encountered in the twenties.
The explanation was required in that time for negative-energy solutions in
the famous equation for spin $j=1/2$.  In the opposite case (if we would
not find a satisfactory interpretation for the $E=0$ solution) we have to
agree with Weinberg~[29b,p.B888]:  ``The fact that these field equations
are of first order for any spin seems to me to be of no great
significance..."} On the other hand, the $m\rightarrow 0$ limit of the
Weinberg equation [and the Dirac-like modification of the Weinberg
equation~\cite{Ahl1}] of the $2j$-order in derivatives is free from all
kinematic acausalities.

\item
M. Moshinsky and A. del Sol Mesa, ref.~\cite{Mosh}, noted
that for a two-equal-mass-particle system
in relativistic quantum mechanics
three possible dispersion relations (Eq. (1.5) of the cited work) exist:
\begin{eqnarray}
E=E_1 +E_2 =\cases{+(2 {\bf p}^2 c^2 + 4m^2 c^4)^{1/2} &\cr
0 &\cr
- (2{\bf p}^2 c^2 + 4m^2 c^4)^{1/2} \quad,&\cr}
\end{eqnarray}
where  ${\bf p_1} = - {\bf p_2} \equiv \frac{1}{\sqrt{2}} {\bf p}$.
They called the infinitely degenerate
level $E=0$ as the ``relativistic cockroach nest",
``as it appears inert so long as the particles are free, but any
interaction immediately brings levels at values $E\neq 0$. The name
[they] have given to these levels comes from its analogy with
a crack in a wall, which looks innocuous under normal circumstances,
but if food is put near it, {\it i.e.} in our case interaction,
the cockroaches start to come out... Analogy indicates
that food near the crack will keep the cockroaches there, and thus
will not hamper our use of the rest of the wall."

\item
In ref.~\cite{Ahl1} an explicit construct of
the  Bargmann-Wightman-Wigner-type quantum field theory
has been proposed in the $(1,0)\oplus (0,1)$ representation space
of the Lorentz group. The remarkable feature of this class of
theories, envisaged in the sixties~\cite{Wigner},
is: a boson and its antiboson can possess opposite relative intrinsic
parities.  The essential ingredients of this model are: the choice of the
spinorial basis (or, rather, of bivectors in the zero-momentum frame) in
the same manner like in the spin $j=1/2$ case:\footnote{See also my recent
works~\cite{Dv-sb}.} $\phi_{_R} (\overcirc{p}^\mu) = \pm \phi_{_L}
(\overcirc{p}^\mu)$; and the use of the Wigner rules~\cite{Wigner2} for
Lorentz transformations of the right- $(j,0)$ and left- $(0,j)$ handed
spinors from the ``rest":
\begin{mathletters} \begin{eqnarray}
(j,0)\,:&&\phi_{_R}
(p^\mu) \,= \,\exp (+\,{\bf J}\cdot\bbox{\varphi})\phi_{_R}
(\overcirc{p}^\mu) \quad,\\ (0,j)\,:&&\phi_{_L} (p^\mu) \,= \,\exp
(-\,{\bf J}\cdot\bbox{\varphi})\phi_{_L} (\overcirc{p}^\mu) \quad.
\end{eqnarray}
\end{mathletters}
As a result, the Weinberg $(1,0)\oplus (0,1)$
configuration-space-free-equation has been
modified:\footnote{Matrices $\gamma_{\mu\nu}$ are
the Barut-Muzinich-Williams covariantly defined matrices~\cite{BarMu}
in the $(1,0)\oplus (0,1)$ representation space.}
\begin{equation}
\left (\gamma_{\mu\nu} \partial^\mu \partial^\nu + \wp_{u,v} m^2\right )
\psi (x) = 0
\end{equation}
with $\wp_{u}=1$ for positive-energy solutions and $\wp_{v}=-1$,
for the negative-energy solutions.

Various problems of describing the particle world in the $(j,0)\oplus
(0,j)$ representation space have also been discussed in the
works~\cite{D40,Ahl-pr,Old}.

\item
In ref.~\cite{D1} the equivalence of the Weinberg field, which satisfies
the equations
\begin{mathletters}
\begin{eqnarray}
(\gamma_{\alpha\beta} p_\alpha p_\beta + m^2 )\psi_1 (x) \,&=& \,0\quad,
\label{eq:a1}\\
(\gamma_{\alpha\beta} p_\alpha p_\beta - m^2) \psi_2 (x)\, &=& \,0\quad,
\label{eq:a2}
\end{eqnarray}
\end{mathletters}
and the antisymmetric tensor field, which satisfies the equations
\begin{mathletters}
\begin{eqnarray}
m^2 F_{\mu\nu} &=&\partial_\mu \partial_\alpha F_{\alpha\nu}
-\partial_\nu \partial_\alpha F_{\alpha\mu} + {1\over 2}
(m^2 - \partial_\lambda^2) F_{\mu\nu}\quad, \label{eq:a3}\\
m^2 \widetilde F_{\mu\nu} &=& \partial_\mu \partial_\alpha \widetilde
F_{\alpha\nu}
-\partial_\nu \partial_\alpha \widetilde F_{\alpha\mu} +
{1\over 2} (m^2 -\partial_\lambda^2) \widetilde F_{\mu\nu}\quad,
\label{eq:a4}
\end{eqnarray}
\end{mathletters}
has been proved.  The concept of the Weinberg field as
a system of two field function $(\psi_1 (x), \quad \psi_2 (x))$
(or  $(F_{\mu\nu} (x),\quad \widetilde F_{\mu\nu} (x))$) has been proposed.
In ref.~\cite{D3} the Weinberg propagators for the spin $j=1$
field have been constructed on the basis of the use
of this set of field functions and parity-conjugates to them.

Let me still note, the equations
(\ref{eq:a1})-(\ref{eq:a4}) have both solutions with
a correct physical dispersion $E^2 -\vec p^{\,2} =m^2$
and tachyonic solutions. The Hammer-Tucker
equation, ref.~\cite{Tucker}, or the Proca equations for
an antisymmetric tensor field:
\begin{mathletters}
\begin{eqnarray}
m^2  F_{\mu\nu} \,&=&\, \partial_\mu \partial_\alpha F_{\alpha\nu}
-\partial_\nu \partial_\alpha F_{\alpha\mu}\quad, \label{eqs} \\
\partial_\lambda^2  F_{\mu\nu} \,&=&\, m^2 F_{\mu\nu}\quad. \label{eqs1}
\end{eqnarray}
\end{mathletters}
(and dual to them) was mentioned in ref.~\cite{D1} to possess six causal
solutions for massive particles. Moreover, in the case $m=0$ the
determinant of the Hammer-Tucker equations is identically equal to
zero.\footnote{In this work we are not going to discuss advantages and
shortcomings coming from the choice of Weinberg or Tucker-Hammer
equations. These issues have been mentioned in previous works but still
deserve a separate paper.} The corresponding Green's function for the
general case (see the similar expression of the Green's function for
massless field in~\cite{Avdeev}), written by using the Euclidean metric, is
\begin{eqnarray}
\lefteqn{G_{\mu\nu,\alpha\beta} (p) = - {1\over c_1 p^2 +c_2 m^2}
\left [\left (\delta_{\mu\alpha}\delta_{\nu\beta} - \delta_{\mu\beta}
\delta_{\nu\alpha}\right ) -\right .\nonumber}\\ &-&\left. {1\over
(c_1+1) p^2 + c_2 m^2} \left ( p_\mu p_\alpha \delta_{\nu\beta} - p_\mu
p_\beta \delta_{\nu\alpha} -p_\nu p_\alpha \delta_{\mu\beta} + p_\nu
p_\beta \delta_{\mu\alpha}\right )\right ]\quad.\label{G}
\end{eqnarray}
The constants $c_{1,2}$ are defined by the equation considered
(see~\cite[Eq.(9)]{D1}). In the case of the Hammer-Tucker equation
$c_1 =0$ and the Green's function is not well-defined in the massless
limit.

\item
In the papers, refs.~\cite{D2,D96}, the transversality (including a
massless case) of the Weinberg (or antisymmetric tensor) fields has been
proved.  This statement opposes to the conclusions of
refs.~\cite{Hayashi,Kalb,Avdeev} and of my previous paper~[47b].
In those papers it was claimed that the quantized massless
antisymmetric tensor field is longitudinal.
What are the origins of such a surprising result
achieved in earlier articles? As a matter of fact it contradicts with
the classical formalism for antisymmetric tensor field
({\it i.e.}, with the Correspondence
Principle) and with the famous Weinberg theorem $B-A=\lambda$,
which allows the helicity of the $(1,0)\oplus (0,1)$ massless field
to be $\lambda =\pm 1$. In  the old works  several authors
have applied the ``generalized Lorentz condition"
imposed on the state vectors (see formulas (18) in ref.~\cite{Hayashi}).
It is the ``generalized Lorentz condition", which coincides with
the equations (\ref{Max1},\ref{Max2}),
that yields the  puzzled dispersion  relation of
the $j=1$ massless field ({\it cf.} with item \# 1).The
papers~\cite{D2,D96} reveal that the use of the well-known equations
(\ref{Max1},\ref{Max2}) in the coordinate space may lead to
confusions such as equating the spin operator to zero.
The result which is
related with our conclusion has been confirmed by
Evans~\cite{Evans1} from different standpoints.  He discussed origins of
appearance of the $B(3)$ longitudinal field, which may be used to obtain
helicities $\pm 1$. Unfortunately,
formal calculations in Evans' papers deserve careful examination.
In the paper~\cite{Avdeev} Avdeev and Chizhov analyzed this problem and
used the Lagrangian which is similar to our Lagrangian except for the
total derivative.\footnote{Our term, ref.~\cite[Eq. (3)]{D96},
$\sim\partial_\mu F_{\nu\alpha} \partial_\nu F_{\mu\alpha}$ is then
equivalent to the additional $-(1/2) (\partial_\mu T_{\mu\alpha})^2$, {\it
cf.} ref.~[51b, Eq.(1,2)]
if one takes into account the possibility of adding total derivatives
to the Lagrangian.}  But, they concerned with the real part of the
antisymmetric tensor field only and lost the information about
possible existence of the $j=1$ antiparticle.  Furthermore,
in the private communications Prof.  L. Avdeev writes himself:  ``In
perturbation theory we systematically ignore any boundary effects,
although they may be of consequence nonperturbatively." Interesting
discussion of the above-mentioned issues can be found in
ref.~\cite{Vlasov}.

Thus,  in the works~\cite{Evans1,D2,D96}
the contradiction between the ``longitudinality"
of the antisymmetric tensor field after quantization and
the Weinberg theorem $B-A=\lambda$  has been partly clarified.

\item
D. V. Ahluwalia {\it et al.} developed the Majorana-McLennan-Case
theory of self/anti-self conjugate states, {\it i.~e.} of
truly neutral particles,  ref.~\cite{DVA1,DVA2}. Complete sets of
second-type spinors
\begin{eqnarray}
\lambda^{S,A} (p^\mu) \, \equiv \, \pmatrix{\left (\zeta_\lambda^{S,A}
\Theta_{[j]}\right ) \,\phi_{_L}^* (p^\mu)\cr
\phi_{_L} (p^\mu)\cr}\quad, \quad
\rho^{S,A} (p^\mu)\, \equiv \, \pmatrix{\phi_{_R} (p^\mu)\cr
\left (\zeta_\rho^{S,A} \Theta_{[j]}\right )^* \,\phi_{_R}^* (p^\mu)\cr}
\end{eqnarray}
have been introduced there. They are {\it not} eigenspinors of the helicity
operator of the $(j,0)\oplus (0,j)$ representation; they describe
the states which are {\it not} eigenstates of the Parity operator;
the self/anti-self charge conjugate states form a {\it bi-orthonormal} set
in a mathematical sense (see~\cite[Eqs.(41-45)]{DVA2} and  my remark in
the footnote \# 2). New fundamental wave equations have been proposed in
the light-front formulation of quantum field theory~\cite{DVA1} and in the
instant form~\cite{DVA2}. As a result of this consideration it was
found that they can describe fermions with the same intrinsic
parities, ref.~\cite{DVOPAR}. The example of such a kind of the theories
has already been discussed in the old paper~\cite{Nigam}, but with type-I
(Dirac) spinors.  Equations for self/anti-self charge conjugate
states recast into the covariant form (the ``MAD" form of the
Dirac equations, see also~\cite{Markov,Ziino,Dv-sb}) in
ref.~\cite{DVO1,DVO10}:  \begin{mathletters} \begin{eqnarray} i \gamma^\mu
\partial_\mu \lambda^S (x) - m \rho^A (x) &=& 0 \quad, \label{11}\\ i
\gamma^\mu \partial_\mu \rho^A (x) - m \lambda^S (x) &=& 0 \quad;
\label{12} \end{eqnarray} \end{mathletters} and \begin{mathletters}
\begin{eqnarray} i \gamma^\mu \partial_\mu \lambda^A (x) + m \rho^S (x)
&=& 0\quad, \label{13}\\ i \gamma^\mu \partial_\mu \rho^S (x) + m
\lambda^A (x) &=& 0\quad.  \label{14} \end{eqnarray} \end{mathletters}
Their possible relevance
to describing neutrino and photon states has been discussed.
Particularly, importance of the axial current has been stressed.

\end{itemize}

On the basis of these thoughts I am  able
to write the {\it ansatzen} substituting the de Broglie
(or Jordan) {\it ansatz}. By the direct examination one can prove
on the classical level
that the 4-vector potential of different polarization states can be
expressed (on using the second-type spinors) as
follows:\footnote{For the sake of the general consideration
it is assumed  the neutrino states to be massive.}
\begin{mathletters}
\begin{eqnarray}
A^\mu ({\bf k}, +1) &=&  {N\over m\sqrt{2}} \left \{
a_1 \left [\overline \lambda^S_\uparrow (k^\mu) \gamma^\mu (1+\gamma^5)
\lambda^S_\downarrow (k^\mu) \right ]  + a_2
\left [\overline \lambda^S_\downarrow (k^\mu) \gamma^\mu (1-\gamma^5)
\lambda^S_\uparrow (k^\mu) \right ]+\right. \, \nonumber\\
&+&  \left . a_3 \left [
\overline \lambda^A_\uparrow (k^\mu) \gamma^\mu (1+\gamma^5)
\lambda^A_\downarrow (k^\mu) \right ] +a_4
\left [\overline \lambda^A_\downarrow (k^\mu) \gamma^\mu (1-\gamma^5)
\lambda^A_\uparrow (k^\mu) \right ]-\right . \, \nonumber\\
&-&  \left . a_5 \left [
\overline \lambda^S_\uparrow (k^\mu) \gamma^\mu (1+\gamma^5)
\lambda^A_\downarrow (k^\mu) \right ] +a_6
\left [\overline \lambda^S_\downarrow (k^\mu) \gamma^\mu (1-\gamma^5)
\lambda^A_\uparrow (k^\mu) \right ]-\right .  \, \nonumber\\
&-&  \left . a_7 \left [
\overline \lambda^A_\uparrow (k^\mu) \gamma^\mu (1+\gamma^5)
\lambda^S_\downarrow (k^\mu) \right ] +a_8
\left [\overline \lambda^A_\downarrow (k^\mu) \gamma^\mu (1-\gamma^5)
\lambda^S_\uparrow (k^\mu) \right ]\right \} \, . \\
A^\mu ({\bf k}, -1) &=& - {N\over m\sqrt{2}}\left\{ b_1 \left [
\overline \lambda^S_\uparrow (k^\mu) \gamma^\mu (1-\gamma^5)
\lambda^S_\downarrow (k^\mu) \right ] +b_2
\left [\overline \lambda^S_\downarrow (k^\mu) \gamma^\mu (1+\gamma^5)
\lambda^S_\uparrow (k^\mu) \right ] +\right .\, \nonumber\\
&+& \left . b_3\left [
\overline \lambda^A_\uparrow (k^\mu) \gamma^\mu (1-\gamma^5)
\lambda^A_\downarrow (k^\mu) \right ] +b_4
\left [\overline \lambda^A_\downarrow (k^\mu) \gamma^\mu (1+\gamma^5)
\lambda^A_\uparrow (k^\mu) \right ] +\right .\, \nonumber\\
&+& \left. b_5 \left [
\overline \lambda^S_\uparrow (k^\mu) \gamma^\mu (1-\gamma^5)
\lambda^A_\downarrow (k^\mu) \right ] -b_6
\left [\overline \lambda^S_\downarrow (k^\mu) \gamma^\mu (1+\gamma^5)
\lambda^A_\uparrow (k^\mu) \right ] +\right .\, \nonumber\\
&+& \left . b_7\left [
\overline \lambda^A_\uparrow (k^\mu) \gamma^\mu (1-\gamma^5)
\lambda^S_\downarrow (k^\mu) \right ] -b_8
\left [\overline \lambda^A_\downarrow (k^\mu) \gamma^\mu (1+\gamma^5)
\lambda^S_\uparrow (k^\mu) \right ] \right \}\, . \\
A^\mu ({\bf k}, 0) &=& {N\over 2m} \left \{ c_1 \left [
\overline \lambda^S_\downarrow (k^\mu) \gamma^\mu
\lambda^S_\downarrow (k^\mu) - \overline \lambda^S_\uparrow
(k^\mu) \gamma^\mu \lambda^S_\uparrow (k^\mu) \right ] +c_2
\left [ \overline \lambda^A _\downarrow (k^\mu) \gamma^\mu
\lambda^A_\downarrow (k^\mu) - \right.\right.\nonumber\\
&-&\left.\left.\overline \lambda^A_\uparrow
(k^\mu) \gamma^\mu \lambda^A_\uparrow (k^\mu) \right ]-
c_3 \left [
\overline \lambda^S_\downarrow (k^\mu) \gamma^\mu \gamma^5
\lambda^A_\downarrow (k^\mu) - \overline \lambda^S_\uparrow
(k^\mu) \gamma^\mu \gamma^5 \lambda^A_\uparrow (k^\mu) \right ] -
\right.\nonumber\\
&-&\left. c_4
\left [ \overline \lambda^A _\downarrow (k^\mu) \gamma^\mu \gamma^5
\lambda^S_\downarrow (k^\mu) - \overline \lambda^A_\uparrow
(k^\mu) \gamma^\mu \gamma^5 \lambda^S_\uparrow (k^\mu) \right ]\right \}
\, .\\
A^\mu ({\bf k}, 0_t) &=& {N\over 2m} \left \{ d_1 \left [
\overline \lambda^S_\downarrow (k^\mu) \gamma^\mu
\lambda^S_\downarrow (k^\mu) + \overline \lambda^S_\uparrow
(k^\mu) \gamma^\mu \lambda^S_\uparrow (k^\mu) \right ] +
d_2 \left [\overline \lambda^A _\downarrow (k^\mu) \gamma^\mu
\lambda^A_\downarrow (k^\mu) + \right.\right.\nonumber\\
&+&\left.\left. \overline \lambda^A_\uparrow
(k^\mu) \gamma^\mu \lambda^A_\uparrow (k^\mu) \right ]
- d_3 \left [
\overline \lambda^S_\downarrow (k^\mu) \gamma^\mu \gamma^5
\lambda^A_\downarrow (k^\mu) + \overline \lambda^S_\uparrow
(k^\mu) \gamma^\mu \gamma^5 \lambda^A_\uparrow (k^\mu) \right ] -
\right.\nonumber\\
&-&\left. d_4 \left [\overline \lambda^A _\downarrow (k^\mu) \gamma^\mu
\gamma^5 \lambda^S_\downarrow (k^\mu) + \overline \lambda^A_\uparrow
(k^\mu) \gamma^\mu \gamma^5 \lambda^S_\uparrow (k^\mu) \right ]\right \}\,.
\end{eqnarray}
\end{mathletters}
Of course, one can repeat the derivation of the formulas of this paper
on using the Dirac 4-spinors, thus arriving at the {\it electron-positron
theory of light}. The room of choosing the constants $a_i, b_i, c_i, d_i$
permits us to obtain various types of (anti)commutation relations for the
composite particles. Tensor currents have also been calculated.
They are also composed of the second-type 4-spinors. In my opinion,
resulting expressions of the type $F^{\mu\nu}_\lambda \sim \bar \nu_\eta
(\sigma^{\mu\nu} \pm i \widetilde \sigma^{\mu\nu}) \nu_{\eta^\prime}$
(and its dual conjugates) also permit to construct the neutrino theory of
light.

Finally, in the papers of Barut, {\it e.g.}, ref.~\cite{Barut}
a self-field formulation of quantum
electrodynamics have been proposed. It is based
on the use of the solution
\begin{equation}\label{pot}
{\cal A}_\mu (x) = \int d^4 y D_{\mu\nu} (x-y) j_\nu (y)
\end{equation}
of the coupled Maxwell-Dirac equation
\begin{equation}
\partial_\mu F_{\mu\nu} (x) = e \overline{\Psi} (x) \gamma_\nu \Psi
(x)\quad.
\end{equation}
$D_{\mu\nu} (x-y)$  is a Green's function of
electromagnetic field in the usual potential formulation.
In a series of the works A. Barut {\it et al.} have shown that this
formulation of quantum electrodynamics (based on the iteration
procedure, {\it not} on the perturbation theory) leads to the same
experimental predictions as the ordinary formalism.

Let me try to write the formula  (\ref{pot}) in the momentum space
(To my knowledge, there were no such attempts in the literature).
I consider  momenta as $q=\lambda t$ and $p= (\lambda -1) t$,
$\lambda$ is some function spanned from $0$  to $1$.
In this case\footnote{It would be interesting to search physical
consequences of the particular choice $\lambda=\frac{m_1}{m_1+m_2}$.}
\begin{eqnarray}
{\cal A}_\mu (x) &=& -e\int {d^3 {\bf p} d^3 {\bf q} \over
(2\pi)^6} \frac{D_F ((q-p)^2)}{2m \sqrt{E_p E_q}}
\sum_{\sigma\sigma^\prime}\left \{ \overline u_\sigma ({\bf p})
\gamma^\mu u_{\sigma^\prime} ({\bf q}) e^{i (q-p) x} a^\dagger_\sigma
({\bf p}) a_{\sigma^\prime} ({\bf q})+ \right .\nonumber\\
&+&\left . \overline v_\sigma ({\bf p}) \gamma^\mu v_{\sigma^\prime}
({\bf q}) e^{-i (q-p) x} b_\sigma ({\bf p})
b^\dagger_{\sigma^\prime} ({\bf q})\right \} \quad,
\end{eqnarray}
and, hence,
\begin{equation}
{\cal A}_\mu (t) = \int_0^1 d\lambda f (\lambda,
t^2 ) \sum_{\sigma\sigma^\prime,\,\pm} \overline\psi^{\pm}_\sigma
((\lambda -1)t) \gamma^\mu \psi^{\pm}_{\sigma^\prime} (\lambda t) \quad.
\end{equation}
Surprisingly, you may see the
well-known Jordan {\it ansatz}.  Thus, referring to the remark of the
previous paragraph one can state the longitudinal de Broglie-Jordan-Barut
potential can describe quantumelectrodynamic processes sufficiently good.
In order to obtain transverse components of the 4-vector potential
one should set up the commutation relations for the 4-spinor field,
which are different from those used in the Dirac theory.

\section{Questions}

This paper speculate about several versions of the construction of
the composite particles from states of the $(1/2,0)\oplus (0,1/2)$
representation space. In
fact, it is a continuation of  efforts undertaken in old papers,
refs.~\cite{Sen,Perkins,Bandy2,Strazhev,Weinberg,Sankar,Tucker,Majorana,CaseM,Barut},
and in the recent
ones~\cite{Ahl0,D4,Ahl1,Old,D1,D2,D3,DVA1,DVA2,DVO1,DVO10,Campo}.  I
realize that not all the topics and not all the contradictions have been
considered here. A task of this paper  was mainly to give a direction for
future researches and to  present a basis for forthcoming publications.

Referring to phenomenological consequences of the theory, I would like to
cite some paragraphs from~[17a]: ``\ldots in view of the
neutrino theory of light, photons are likely to interact weakly also,
apart from the usual electromagnetic interactions\ldots This assumed
photon-neutrino weak interaction,
if it exists, will have important bearing on astrophysics.
In fact, this interaction can then be held responsible for
the following neutrino-generating processes in stars:
\begin{mathletters}
\begin{eqnarray}
&&(1) \quad \gamma  + e^- \leftrightarrow e^- +\nu +\bar \nu \quad, \\
&&(2) \quad e^- + Z \leftrightarrow e^- +Z +\nu + \bar \nu \quad, \\
&&(3) \quad e^- +e^ + \leftrightarrow \nu +\bar \nu \quad, \\
&&(4) \quad \gamma + \gamma \leftrightarrow \nu +\bar \nu \quad, \\
&&(5) \quad \gamma  + \gamma \leftrightarrow \gamma+\nu +\bar\nu \quad, \\
&&(6)\quad  \Gamma \rightarrow \nu +\bar\nu  \,
(\Gamma\rightarrow e^- +e^+ \rightarrow \gamma\rightarrow \nu
+\bar \nu) \quad (plasma \,\, process)\quad.
\end{eqnarray}
\end{mathletters}
\ldots The energy dependence of the cross sections for these
processes according to the present
theory will be significantly different from that in other theories."

\bigskip

Next, I would like to put the following problems forth:

\begin{itemize}

\item
In neutrino physics we have now most consistent indications for
new physics\cite{exper,APS}. What are the origins of negative
mass-squared problem from a viewpoint of the presented
model? Can the solar neutrino deficit be caused by the processes mentioned
above (like $\bar \nu \nu \rightarrow 2\gamma$)?  Perhaps, we should
re-calculate the old of the Sun on the basis of this model?

\item
Can recently observed neutrino oscillations  (for a reviews see
ref.~\cite{oscil,DVO1}) have some relations with the
presented model?

\item
It is known that the Aharonov-Bohm effect, ref.~\cite{AB},
cannot be explained on the basis of transversal electric and magnetic
fields ${\vec E}$ and ${\vec B}$. Other authors  applied to
4-vector potential in order to explain it. Perhaps, the possibility of
dual solutions $F_{\mu\nu}$ and ${\tilde F_{\mu\nu}}$ for the spin-1
field and/or the existence of Majorana-like spinors have been missed in
previous attempts? Several recent publications indicate the possibility of
explanation of the Aharonov-Bohm effect on the basis of the longitudinal
solutions of electromagnetism~\cite{CHUB2} ({\it cf.} also
ref.~\cite{Whittaker}).

\end{itemize}

We are sure there exist several other problems in the modern field theory,
which can be related with the matters discussed in this paper.

\smallskip

{\bf Acknowledgements.}  I thank Prof. D. V. Ahluwalia
whose papers helped me and who gave an initial impulse
to come back  to these topics. I greatly appreciate communications
with  A.~F.~Pashkov several years ago. His remarks have been definitely
helpful to realise new perspectives in the old theory. Perhaps, the
presented results have  been obtained (but, unfortunately, not published)
by him several years ago.

I am grateful to Zacatecas University for a professorship.

\end{document}